\documentclass[onecolumn,preprintnumbers,eqsecnum,12pt]{revtex4}
\usepackage{amsfonts}
\usepackage{amssymb}
\usepackage{amsmath}
\usepackage{epsf}
\usepackage{graphicx}
\usepackage{dcolumn}
\usepackage{bm}

\begin{document}

\title{Hawking radiation from the cosmological horizon in a FRW universe}
\author{Ya-Peng Hu}
\email{yapenghu@itp.ac.cn}

\address{
Center for High Energy Physics, Peking University, Beijing 100871, China}
\address{
Key Laboratory of Frontiers in Theoretical Physics, Institute of
Theoretical Physics, Chinese Academy of Sciences, P.O. Box 2735,
Beijing 100190, China}

\begin{abstract}
It is well known that there is a Hawking radiation from the cosmological
horizon of the de-sitter spacetime, and the de-sitter spacetime can be a
special case of a FRW universe. Therefore, there may be a corresponding
Hawking radiation in a FRW universe. Indeed, there have been several clues
showing that there is a Hawking radiation from the apparent horizon of a FRW
universe. In our paper, however, we find that the Hawking radiation
may come from the cosmological horizon. Moreover, we also find
that the Hawking radiation from the apparent horizon of a FRW universe in
some previous works can be a special case in our result, and the condition is that the
variation rate of cosmological horizon $\overset{.}{r}_{H}$ is zero. Note that, this condition is also consistent with the underlying integrable condition in these works from the apparent horizon.

\end{abstract}

\maketitle

\vspace*{1.cm}

\newpage

\section{Introduction}

Since Hawking found that there was a thermal radiation like a black body in
a black hole, it has been further found that the radiation is in fact due to
the existence of event horizon~\cite{Hawking:1974sw}. The event horizon played a key point can
also be seen from the Unruh effect where an uniformly accelerating observer
with acceleration $a$ in the Minkowskian spacetime can detect a thermal
spectrum with temperature $T=a/2\pi$~\cite{Unruh:1976db}. Here the Unruh radiation is closely
related to the existence of Rindler causal horizon for the observer.
Obviously, the Hawking radiation with a temperature which is proportional to its surface
gravity on the event horizon can give some insight on the deep relationship
between gravity and thermodynamics. Indeed, the thermodynamics of black hole has already been constructed
with the Bekenstein entropy of a black hole~\cite{Bekenstein:1972tm,Bekenstein:1973mi,Bekenstein:1973ur,Bardeen:1973gs}. Note
that, the Hawking radiation is usually investigated from the event horizon of a
stationary black hole. In fact, it can also be obtained from the cosmological
horizon of a spacetime such as the cosmological horizon of de Sitter
spacetime~\cite{GH,Parikh}.

Event horizon and cosmological horizon are both global concepts~\cite{Hawking:1973uf,Wald:1984rg}.
Strictly speaking, locally it is not known whether there is an event
horizon or cosmological horizon associated with a certain dynamical
spacetime at some time. Thus this causes the difficulty to discuss Hawking radiation for a dynamic spacetime.
However, by using the the null property of event horizon or cosmological horizon and
the intrinsic symmetry of a dynamic spacetime, we can find a corresponding hypersurface which can
reduce the event horizon or cosmological horizon in the stationary case. Because of this,
we also call this corresponding hypersurface as the event horizon or cosmological horizon for a dynamic spacetime in our paper~\cite{Li:2000rk,Hiscock:1977qe,Zhao:1992ad,Li:1999xz,Cai:2008mh}.
In spite of that, another situation appears. This is, the event horizon (the above corresponding hypersurface) and apparent horizon for a dynamic spacetime are usually different, while they are consistent for a stationary spacetime. Therefore, the Hawking radiation from
which horizon is still an open question. Recently, Hayward and other authors
have attacked this question~\cite{Cai:2008mh,Nielsen:2008cr,Hay}. By using the quasi-local Misner-Sharp energy~\cite{Misner:1964je,Maeda:2007uu,Cai:2009qf}, the so-called unified first law can
be deduced from the Einstein equation in a spherical symmetric
spacetime~\cite{Hayward:1993wb,Hayward:1994bu,Hayward:1997jp,Hayward:1998ee}. And they argued that the Hawking radiation might come from the
apparent horizon for a dynamic spherical symmetric black hole spacetime,
since after projecting the unified first law on the apparent horizon of a
dynamic spherical symmetric black hole spacetime, one can obtain an analogy
of the first law of thermodynamics of stationary black hole. In addition,
one could use the Hamilton-Jacobi equation of particles to make a simple
proof~\cite{Hay,Padd}. However, there are other works showing that
the Hawking radiation can come from the event horizon of dynamic black hole
spacetime by investigating the behavior of the quantum filed near the event
horizon~\cite{Li:2000rk,Hiscock:1977qe,Zhao:1992ad,Li:1999xz}.

On the other hand, the Friedmann-Robertson-Walker (FRW) universe is a dynamical spacetime, and the de
Sitter spacetime can be its special case. Therefore, Hawking radiation may also exist in a FRW universe. By considering that the FRW universe is
also a spherical symmetric spacetime and with an apparent horizon, therefore, the
above discussion on the apparent horizon of dynamic spherical symmetric
black hole spacetime can be generalized to the FRW universe. There have been many interesting works based on this issue ~\cite{CC1,AC2,CC2,Cai,Ge,Cai:2008gw,Li:2008gf,Zhu:2009wa}, and it has been proved that
the Hawking temperature of the apparent horizon in a FRW universe is $%
T=1/2\pi r_A$, where the temperature is measured by the corresponding Kodama
observer~\cite{Kodama:1979vn} and $r_A$ is the radius of apparent horizon~\cite{Cai:2008gw,Li:2008gf}. In particular, note here that if we assume the entropy of apparent horizon $S$
satisfying $S= A/4$, where $A$ is the area of the apparent horizon, one is
able to derive Friedmann equations of the FRW universe with any spatial
curvature by applying the Clausius relation to apparent horizon~\cite{Cai:2005ra,Cai:2008ys}. However,
there is the same situation as the dynamic black hole spactime that the
cosmological horizon of a FRW universe is not usually consistent with its
apparent horizon. Therefore, one of our motivations is that which kind of results we will obtain if we investigate the
behavior of the quantum filed near the cosmological horizon of FRW universe.

There are several methods to investigate the behavior of quantum filed near
the horizon of a spacetime~\cite{Kokkotas:1999bd,Robinson:2005pd,Damour:1976jd}. In our paper, we mainly use the Damour-Ruffini method first
proposed by Damour and Ruffini and then developed by Sannan and Zhao~\cite{Damour:1976jd,Zhao:1992ad,Li:1999xz,Hu:2006ct}. By
using the fact that usually the Klein-Gordon equation in the tortoise
coordinates can be reduced to the standard form of wave equation near the
cosmological horizon of FRW universe, we obtain the appropriate
parameter $\kappa$ which corresponds to the surface gravity in the
stationary case. Moreover, we find that the ingoing wave of FRW universe
is not analytical on the cosmological horizon, and it can be extended by
analytical continuation from the inside to outside of the cosmological horizon ~\cite{Damour:1976jd,Penrose:1968me,Zhao:1992ad,Li:1999xz,Hu:2006ct}. After doing these, we obtain the Hawking radiation with the
temperature on the cosmological horizon of a FRW universe.




The organization of the paper is as follows. In Sec.~II, we first obtain the
cosmological horizon in a FRW universe, and then use the Damour-
Ruffini method to obtain its Hawking radiation. Sec.~III is devoted to the
conclusion and discussion.

\section{The cosmological horizon and its Hawking radiation in a FRW universe}

The metric of a FRW universe is
\begin{equation}
ds^{2}=-dt^{2}+a^{2}(t)\left( \frac{d\rho ^{2}}{1-k\rho ^{2}}+\rho
^{2}d\Omega _{2}^{2}\right) ,  \label{FRWmetric1}
\end{equation}%
where $t$ is the cosmic time, $\rho $ is the comoving radial coordinate, $a$
is the scale factor, $d\Omega _{2}^{2}$ denotes the line element of a $2$%
-dimensional sphere with unit radius, $k=1$, $0$ and $-1$ represent a
closed, flat and open FRW universe respectively.

For the convenience, we define $r=a\rho $. Therefore, the metric (\ref%
{FRWmetric1}) can be rewritten as
\begin{equation}
ds^{2}=-\frac{1-r^{2}/r_{A}^{2}}{1-kr^{2}/a^{2}}dt^{2}-\frac{2Hr}{%
1-kr^{2}/a^{2}}dtdr+\frac{1}{1-kr^{2}/a^{2}}dr^{2}+r^{2}d\Omega _{2}^{2},
\label{FRWmetric2}
\end{equation}%
where $r_{A}=1/\sqrt{H^{2}+k/a^{2}}$ is the location of apparent horizon in
a FRW universe.

Note that the metric of the de Sitter spacetime is
\begin{equation}
ds^{2}=-\left( 1-\frac{r^{2}}{l^{2}}\right) dt^{2}+\left( 1-\frac{r^{2}}{%
l^{2}}\right) ^{-1}dr^{2}+r^{2}d\Omega _{2}^{2}.  \label{ineq1}
\end{equation}%
and the FRW metric (\ref{FRWmetric2}) can become
\begin{equation}
ds^{2}=-\frac{1-r^{2}/r_{A}^{2}}{1-kr^{2}/a^{2}}(dt+\frac{Hr}{%
1-r^{2}/r_{A}^{2}}dr)^{2}+\frac{1}{1-r^{2}/r_{A}^{2}}%
dr^{2}+r^{2}d\Omega _{2}^{2}.  \label{FRWmetric3}
\end{equation}%
Therefore, it can be easily found that the de Sitter spacetime is just a special
case of the FRW universe where $k=0$ and $r_{A}=H^{-1}=l$ is a constant in (%
\ref{FRWmetric3}). On the other hand, we know that $r=l$ is the cosmological
horizon of the de Sitter spacetime, therefore, there may be a corresponding
cosmological horizon in a FRW universe. By using the null property of the
cosmological horizon and the spherical symmetry in (\ref{FRWmetric2}), we
can indeed obtain that the corresponding cosmological horizon $r=r_{H}(t)$
which satisfies
\begin{equation}
g^{\mu \nu }\frac{\partial f}{\partial x^{\mu }}\frac{\partial f}{\partial
x^{\nu }}=0,
\end{equation}%
is
\begin{equation}
1-r_{H}^{2}/r_{A}^{2}=\overset{.}{r}_{H}^{2}-2Hr_{H}\overset{.}{r}_{H}.
\label{CosmologyHorizon}
\end{equation}%
where $f=r-r_{H}(t)$. From (\ref{CosmologyHorizon}), it can be also easily
checked that the corresponding cosmological horizon $r_{H}(t)$ is just the
cosmological horizon of the de Sitter spacetime when $k=0$ and $\overset{.}{r%
}_{H}=0.$

In the following, we investigate the Hawking temperature of the
corresponding cosmological horizon $r=r_{H}(t)$ in a FRW universe. For the
simplicity, we just consider the Klein-Gordon field in a FRW universe. And
the Klein-Gordon equation
\begin{equation}
(\square -m^{2})\Phi =\frac{1}{\sqrt{-g}}\frac{\partial }{\partial x^{\mu }}(%
\sqrt{-g}g^{\mu \nu }\frac{\partial }{\partial x^{v}})\Phi -m^{2}\Phi =0.
\label{KGEQ}
\end{equation}%
can be rewritten in the FRW coordinates (\ref{FRWmetric2}) such that
\begin{align}
& -\frac{\partial }{\partial t}(\frac{1}{\sqrt{1-\frac{k}{a^{2}}r^{2}}}\frac{%
\partial }{\partial t})\frac{\rho (t,r)}{r}-\frac{\partial }{\partial t}(%
\frac{Hr}{\sqrt{1-\frac{k}{a^{2}}r^{2}}}\frac{\partial }{\partial r})\frac{%
\rho (t,r)}{r}-\frac{1}{r^{2}}\frac{\partial }{\partial r}(\frac{r^{2}}{%
\sqrt{1-\frac{k}{a^{2}}r^{2}}}Hr\frac{\partial }{\partial t})\frac{\rho (t,r)%
}{r}  \notag \\
& +\frac{1}{r^{2}}\frac{\partial }{\partial r}[\frac{r^{2}}{\sqrt{1-\frac{k}{%
a^{2}}r^{2}}}(1-r^{2}/r_{A}^{2})\frac{\partial }{\partial r}]\frac{\rho (t,r)%
}{r}=[m^{2}+\frac{l(l+1)}{r^{2}}]\frac{1}{\sqrt{1-\frac{k}{a^{2}}r^{2}}}%
\frac{\rho (t,r)}{r},  \label{RadialEq}
\end{align}%
\begin{equation}
\frac{1}{\sin \theta }\frac{\partial }{\partial \theta }(\sin \theta \frac{%
\partial }{\partial \theta })Y_{lm}(\theta ,\varphi )+\frac{1}{\sin
^{2}\theta }\frac{\partial ^{2}}{\partial \varphi ^{2}}Y_{lm}(\theta
,\varphi )+l(l+1)Y_{lm}(\theta ,\varphi )=0,
\end{equation}%
where $m$ is the rest mass of the Klein-Gordon particle, $Y_{lm}(\theta
,\varphi )$ is the usual spherical harmonics and $\Phi $ has been separated
as
\begin{equation}
\Phi =\frac{1}{r}\rho (t,r)Y_{lm}(\theta ,\varphi ).
\label{FunctionSeperated}
\end{equation}%
In order to investigate the behavior of the scalar field near the
cosmological horizon, we introduce the generalized tortoise coordinate
transformation
\begin{eqnarray}
r_{\ast } &=&r+\frac{1}{2\kappa }\ln [r_{H}(t)-r],  \notag \\
t_{\ast } &=&t-t_{0},  \label{Transforamtion}
\end{eqnarray}%
where $\kappa $ is an adjustable constant, $r_{H}(t)$ is just the
location of the cosmological horizon, and $t_{0}$ is a constant representing the time when the particles are radiated from the horizon. Note that, $\kappa $ can be just the
surface gravity of the event horizon or cosmological horizon in the
stationary spacetimes.

From (\ref{Transforamtion}), the radial equation (\ref%
{RadialEq}) becomes
\begin{align}
& \{-\frac{2\kappa (r-r_{H})(\overset{.}{a}^{2}+k+a\overset{..}{a})}{%
a[r(2r\kappa -2r_{H}\kappa +1)\overset{.}{a}-a\overset{.}{r}_{H}]}+\frac{%
2\left( l^{2}+l+m^{2}r^{2}\right) \kappa a(r-r_{H})}{r^{2}[r(2r\kappa
-2r_{H}\kappa +1)\overset{.}{a}-a\overset{.}{r}_{H}]}\}\rho  \notag \\
& +\{\frac{-[\overset{.}{r}_{H}^{2}+(r-r_{H})\overset{..}{r}%
_{H}-1]a^{2}+[(r+r_{H})\overset{.}{a}\overset{.}{r}_{H}+r(r-r_{H})(2r\kappa
-2r_{H}\kappa +1)\overset{..}{a}]a}{a(r-r_{H})[r(2r\kappa -2r_{H}\kappa +1)%
\overset{.}{a}-a\overset{.}{r}_{H}]}  \notag \\
& +\frac{r[2\kappa r^{2}+2\kappa r_{H}^{2}-(4r\kappa +1)r_{H}](\overset{.}{a}%
^{2}+k)}{a(r-r_{H})[r(2r\kappa -2r_{H}\kappa +1)\overset{.}{a}-a\overset{.}{r%
}_{H}]}\}\frac{\partial \rho }{\partial r_{\ast }}+\{\frac{(2r\kappa
-2r_{H}\kappa +1)^{2}(\overset{.}{a}^{2}+k)r^{2}}{2\kappa
a(r-r_{H})[r(2r\kappa -2r_{H}\kappa +1)\overset{.}{a}-a\overset{.}{r}_{H}]}
\notag \\
& +\frac{-2(2r\kappa -2r_{H}\kappa +1)\overset{.}{a}\overset{.}{r}%
_{H}ar+a^{2}[-(2r\kappa +1)^{2}+4\kappa r_{H}(2r\kappa +1)-4\kappa
^{2}r_{H}^{2}+\overset{.}{r}_{H}^{2}]}{2\kappa a(r-r_{H})[r(2r\kappa
-2r_{H}\kappa +1)\overset{.}{a}-a\overset{.}{r}_{H}]}\}\frac{\partial
^{2}\rho }{\partial r_{\ast }^{2}}  \notag \\
& +\frac{2\kappa (r-r_{H})\overset{.}{a}}{r(2r\kappa -2r_{H}\kappa +1)%
\overset{.}{a}-a\overset{.}{r}_{H}}\frac{\partial \rho }{\partial t_{\ast }}%
+2\frac{\partial ^{2}\rho }{\partial t_{\ast }\partial r_{\ast }}+\frac{%
2\kappa a(r-r_{H})}{r(2r\kappa -2r_{H}\kappa +1)\overset{.}{a}-a\overset{.}{r%
}_{H}}\frac{\partial ^{2}\rho }{\partial t_{\ast }^{2}}=0.  \label{RadialEq2}
\end{align}%
when $r\rightarrow r_{H}$ and $t\rightarrow t_{0}$, the radial equation (\ref%
{RadialEq2}) is
\begin{equation}
A\frac{\partial ^{2}\rho }{\partial r_{\ast }^{2}}+2\frac{\partial ^{2}\rho
}{\partial t_{\ast }\partial r_{\ast }}+\alpha _{0}\frac{\partial \rho }{%
\partial r_{\ast }}=0,  \label{RadialEq3}
\end{equation}%
where we have used the equation (\ref{CosmologyHorizon}) and
\begin{equation}
A=-\frac{H\overset{.}{r}_{H}-(H^{2}+k/a^{2})r_{H}}{\kappa (Hr_{H}-\overset{.}%
{r}_{H})}+2\overset{.}{r}_{H},~\alpha _{0}=\frac{(H^{2}+k/a^{2})r_{H}-H%
\overset{.}{r}_{H}+\overset{..}{r}_{H}-\frac{\overset{..}{a}}{a}r_{H}}{%
\overset{.}{r}_{H}-Hr_{H}}.
\end{equation}%
The two linearly independent solutions of (\ref{RadialEq3}) are
\begin{equation}
\rho _{out}=e^{-i\omega t_{\ast }},  \label{Solution1} \\
\end{equation}
and
\begin{equation}
\rho _{in}=e^{-i\omega t_{\ast }+2i\omega r_{\ast }/A}e^{-\alpha
_{0}r_{\ast }/A}.  \label{Solution2}
\end{equation}
which is just inside the cosmological horizon ($r<r_{H}$). By using the fact that usually the Klein-Gordon equation in the tortoise
coordinates can be reduced to the standard form of wave equation near the
horizon~\cite{Damour:1976jd,Zhao:1992ad,Li:1999xz,Hu:2006ct}
\begin{equation}
\frac{\partial ^{2}\rho }{\partial r_{\ast }^{2}}+2\frac{\partial ^{2}\rho }{%
\partial t_{\ast }\partial r_{\ast }}=0,  \label{StandardForm}
\end{equation}%
we can adjust the parameter $\kappa$ to make $A=1$, and
\begin{equation}
\kappa =\frac{H\overset{.}{r}_{H}-(H^{2}+k/a^{2})r_{H}}{(Hr_{H}-\overset{.}{r%
}_{H})(2\overset{.}{r}_{H}-1)}.  \label{SurfaceGravity}
\end{equation}%
Note that, $A=1$ can also be implied from the special case, that of the de
Sitter spacetime. In this special case, $k=0$ and $\overset{.}{r}_{H}=0$
with $r_{A}=H^{-1}=l$, the $\kappa$ in (\ref{SurfaceGravity}) is $\kappa=
1/l$ which is just the surface gravity of the cosmological horizon in the
de Sitter spacetime. In addition, from (\ref{SurfaceGravity}),
it can also be found that $\kappa$ is indeed a constant just related to $t_{0}$.

Therefore, the ingoing wave of the Klein-Gordon filed near the cosmological
horizon can be further rewritten as
\begin{equation}
\rho _{in}=Ce^{-i\omega t_{\ast }+2i\omega r_{\ast }}e^{-\alpha _{0}r_{\ast
}}=Ce^{-i\omega t_{\ast }}e^{2i\omega r-\alpha _{0}r}(r_{H}-r)^{i\omega
/\kappa -\alpha _{0}/2\kappa }.  \label{Ingoingmode}
\end{equation}%
where we have used (\ref{Transforamtion}) and added the normalized factor $C$. Note that, $\rho _{out}$ represents an outgoing wave and is well-behaved when analytically extended outside $r>r_{H}$. However, we can find that the ingoing
wave $\rho _{in}$ (\ref{Ingoingmode}) has a logarithmic singularity at the cosmological horizon $r=r_{H}$ and
is not analytical on the cosmological horizon. Thus
we can extend it by analytical continuation from the inside to outside of the cosmological
horizon~\cite{Damour:1976jd,Zhao:1992ad,Li:1999xz,Hu:2006ct,Penrose:1968me}
\begin{equation}
(r_{H}-r)\rightarrow |r_{H}-r|e^{i\pi }=(r-r_{H})e^{i\pi },
\end{equation}%
and then the ingoing wave (\ref{Ingoingmode}) becomes
\begin{equation}
{\tilde \rho }_{in}=Ce^{-i\omega
t_{\ast }}e^{2i\omega r-\alpha _{0}r}(r-r_{H})^{i\omega /\kappa -\alpha
_{0}/2\kappa }e^{-\frac{i\pi \alpha _{0}}{2\kappa }}e^{-\frac{\pi \omega }{%
\kappa }}=Ce^{-i\omega t_{\ast }+2i\omega r_{\ast }}e^{-\alpha _{0}r_{\ast
}}e^{-\frac{i\pi \alpha _{0}}{2\kappa }}e^{-\frac{\pi \omega }{\kappa }%
},r>r_{H}.  \label{Ingoingmode2}
\end{equation}%
By using the Heaviside function $Y$
\begin{equation}
Y(x)=\Big\{_{0,~~~x~<0}^{1,~~~x~\geq 0}
\end{equation}%
the complete ingoing wave can be
\begin{equation}
\phi _{\omega }^{in}=N_{\omega }[Y(r_{H}-r)\rho _{in}+Y(r_{H}-r){\tilde \rho }_{in}].  \label{CompleteIngoingwave}
\end{equation}%
where $N_{w}$ is a normalization factor.
Physically, the waves (\ref{Ingoingmode}) (\ref{Ingoingmode2})
can represent an ingoing particle wave inside the cosmological horizon and an outgoing antiparticle wave of negative energy outside the cosmological horizon~\cite{Damour:1976jd}. Another interpretation is that an antiparticle of positive energy ingoing in the past being scattered forward in time at the cosmological horizon, and the ingoing wave describing this antiparticle state is just (\ref{CompleteIngoingwave}).
Therefore, similar to the WKB approximation in the quantum mechanical barrier penetration, $N_{\omega }^{2}$ can represent the strength of a particle wave ingoing or tunneling from the cosmological horizon. More details, the antiparticle state $\phi _{\omega }^{in}$ is split into two components, a particle wave of strength $N_{\omega }^{2}$ ingoing from the horizon and a negative-energy flux of antiparticles $N_{\omega }^{2}$ outgoing in the future toward the outside of the cosmological horizon. The latter can always be interpreted as an antiparticle wave of strength $N_{\omega }^{2}e^{-2 \pi \omega /\kappa}$  with positive energy flux ingoing in the past from the outside of the cosmological horizon~\cite{Damour:1976jd}.

In the following, we use a simple argument to obtain the temperature of radiation~\cite{Damour:1976jd,Zhao:1992ad,Li:1999xz,Hu:2006ct}. As $\rho_{in}$ is already normalized,
the scalar product of $\phi _{\omega }^{in}$ in (\ref{CompleteIngoingwave}) is
\begin{equation}
(\phi _{\omega _{1}}^{in},\phi _{\omega _{2}}^{in})=N_{\omega _{1}}N_{\omega
_{2}}(\delta _{\omega _{1}\omega _{2}}-e^{-\pi (\omega _{1}+\omega
_{2})/\kappa }\delta _{\omega _{1}\omega _{2}}),  \label{Normalization}
\end{equation}%
where we have used the fact that the inner product of the wave function is normalized to minus $\delta$ function for the Bose particle with negative energy. Note that, if $\kappa <0$ in (\ref{Normalization}), we obtain
\begin{equation}
(\phi _{w}^{in},\phi _{w}^{in})=-1=N_{\omega }^{2}(1-e^{-2\pi \omega /\kappa
}).
\end{equation}%
which is just a thermal spectrum with a temperature $T=-\kappa /2\pi $.
While if $\kappa >0$, we obtain
\begin{equation}
(\phi _{\omega }^{in},\phi _{\omega }^{in})=1=N_{\omega }^{2}(1-e^{-2\pi
\omega /\kappa }).
\end{equation}%
which is apparently not a thermal spectrum. However, we can redefine the
complete ingoing wave in (\ref{CompleteIngoingwave}) as
\begin{equation}
\phi _{\omega }^{in^{\prime }}=e^{\frac{\pi \omega }{\kappa }}N_{\omega
}[Y(r_{H}-r)\rho _{in}+Y(r_{H}-r)\overset{\symbol{126}}{\rho }_{in}],
\end{equation}%
from which we obtain
\begin{equation}
(\phi _{\omega }^{in^{\prime }},\phi _{\omega }^{in\prime })=1=N_{\omega
}^{2}(e^{2\pi \omega /\kappa }-1).
\end{equation}%
which is a thermal spectrum with the temperature $T=\kappa /2\pi $.

In other words, we obtain the thermal spectrum in both cases
\begin{equation}
N_{\omega }^{2}=1/[\exp (\omega /K_{B}T)-1],
\end{equation}%
and the temperature $T$ is
\begin{equation}
T=\frac{|\kappa |}{2\pi }=|\frac{(H^{2}+k/a^{2})r_{H}-H\overset{.}{r}_{H}}{%
2\pi (Hr_{H}-\overset{.}{r}_{H})(2\overset{.}{r}_{H}-1)}|.  \label{Temperture}
\end{equation}

\section{Conclusion and discussion}

Whether there is a Hawking radiation in a FRW universe is a very interesting
question. From the fact that the de Sitter spacetime can be a
special case of a FRW universe and there is a Hawking radiation from the
cosmological horizon of the de Sitter spacetime, therefore, it may also have a
corresponding Hawking radiation in a FRW universe. Indeed, there have been
some clues showing that there is a Hawking radiation from the apparent
horizon in a FRW universe. However, in our paper, after finding the
corresponding cosmological horizon of a FRW universe first, and then investigating
the behavior of a Klein-Gordon field near the cosmological horizon, we
obtain that the Hawking radiation comes from the cosmological horizon of a
FRW universe. Note that, when $\overset{.}{r}_{H}=0$, we can see that the cosmological horizon in (\ref{CosmologyHorizon}) is same with the apparent horizon. And the temperature in (\ref{Temperture}) is
\begin{equation}
T=\frac{1}{2\pi Hr_{A}^{2}},  \label{TemperatureSpecial1}
\end{equation}%
which is apparently not same as the temperature $T=\frac{1}{2\pi r_{A}}$ in some previous results from the apparent horizon~\cite{Cai:2008gw,Li:2008gf} . However, there the temperature $T=\frac{1}{2\pi r_{A}}$ is measured by the Kodama observer. From which, the temperature measured by the observer $(\partial / \partial t)^{a} $ in (\ref{FRWmetric2}) is $T=\frac{1}{2\pi Hr_{A}^{2}}$~\cite{Cai:2008gw}. Furthermore, $\overset{.}{r}_{H}=0$ ensures
the observer in the coordinates system in (\ref{Transforamtion}) same as the
observer $(\partial / \partial t)^{a} $ in (\ref{FRWmetric2}). Therefore, our
result under the condition $\overset{.}{r}_{H}=0$ is in fact consistent with the result in reference~\cite{Cai:2008gw,Li:2008gf}. In addition, we can further find that this condition $\overset{.}{r}_{H}=0$ is consistent with the underlying integrable condition in~\cite{Cai:2008gw,Li:2008gf}. From $\overset{.}{r}_{H}=0$, we can find that the cosmological horizon and apparent horizon are same. Therefore, $\overset{.}{r}_{H}=0$ can reduce $\overset{.}{r}_{A}=0$. On the other hand, from equations (7) and (10) in~\cite{Cai:2008gw}, the underlying integrable condition coming from $\partial_{\tilde r} \partial_{t} S=\partial_{t}\partial_{\tilde r}S$ can also deduce $\overset{.}{r}_{A}=0$. All these consistences partly support the validity of our temperature (\ref{Temperture}).

It should be emphasized that our temperature (\ref{Temperture}) is also valid just under some conditions (like the quasi-static or adiabatic condition), which can be implicated from the limits $r\rightarrow r_{H}$ and $t\rightarrow t_{0}$ in (\ref{RadialEq3}). Therefore, it would be very interesting to have further research on the validity of this temperature (\ref{Temperture}) to obtain the explicit conditions. In addition, the temperature from the apparent horizon is obtained by using the Hamilton-Jacobi equation, and the Hamilton-Jabobi equation can be a WKB approximation solution of the Klein-Gordon equation. Thus it is also an interesting work to research the approximation condition from the Klein-Gordon equation to the Hamilton-Jabobi equation.
Moreover, from the modern quantum field theory, the temperature comes from two different vacuums~\cite{Birrell:1982ix}, and there have been some works to show the two corresponding different vacuums in the Hamilton-Jacobi method, therefore, the two underling different vacuums in the Damour-Ruffini method are also interesting to find out~\cite{Majhi,Zhao:1994dt,FurtherPaper}. In addition, it should be noted that there is a different approach of particle production in a FRW universe named Parker particle production~\cite{Parker:1968mv}. Apparently, the particle production in our paper is different from the Parker particle production for two reasons. First, the spectrums of numbers of particles in Parker particle production are usually not an absolutely thermal spectrum of black body. Second, the two different vacuums constructed are apparently different~\cite{Majhi,Zhao:1994dt,FurtherPaper}. Therefore, it would be very interesting to give further study on the underlying relationship between the particle production in our paper and the Parker particle production.


\section{Acknowledgements}

Y.P Hu thanks Professors Rong-Gen Cai, Zheng Zhao and Gui-Hua Tian for their helpful discussions. Y.P Hu also thanks very much Professor D.Singleton for his communications.
This work is supported partially by grants from NSFC, China (No. 10773002, No. 10875018, No. 10873003
and No. 10975168).



\end{document}